\documentclass[fleqn,usenatbib]{mnras}

\usepackage[T1]{fontenc}

\DeclareRobustCommand{\VAN}[3]{#2}
\let\VANthebibliography\thebibliography
\def\thebibliography{\DeclareRobustCommand{\VAN}[3]{##3}\VANthebibliography}

\usepackage{graphicx}	% Including figure files
\graphicspath{{figures/}}
\usepackage{amsmath}	% Advanced maths commands
\usepackage{amssymb}
\usepackage{booktabs}
\DeclareMathAlphabet\mathbfcal{OMS}{cmsy}{b}{n}
\title[Bars and Warps in Rotating Haloes]{The Formation of Bars and Warps in Rotating Haloes}
\author[R. Joshi et al.]{
Robin Joshi,\thanks{E-mail: 16rj17@queensu.ca}
Lawrence M. Widrow,\thanks{E-mail: widrow@queensu.ca}
\\
Department of Physics, Engineering Physics and Astronomy, Queen's University, Kingston, K7L 3X5, Canada
}

\date{Accepted XXX. Received YYY; in original form ZZZ}

\pubyear{2023}

\begin{document}
\label{firstpage}
\pagerange{\pageref{firstpage}--\pageref{lastpage}}
\maketitle

\begin{abstract}
We investigate the effects of halo kinematics on the dynamics of stellar discs by simulating the evolution of isolated disc-halo systems from equilibrium initial conditions. Our main results come from four simulations where the initial disc is identical and the halo is either treated as a rigid potential or is "live" with isotropic orbits or orbits that preferentially rotate with or counter to the disc. We confirm previous results that bar formation is more vigorous in models with a live halo than a rigid one and is further enhanced when halo orbits preferentially rotate with the disc. We discuss two types of buckling events with different symmetries about the mid plane, one that occurs just as the bar is forming and the other well after the bar has been established. We also show that warps are most easily excited and maintained when the halo is counter-rotating with the disc, in agreement with theoretical predictions. Our most novel result is the discovery of a rotating halo instability, which causes the disc and halo cusp to spiral outward from the center of mass of the system whether the halo rotates with the disc or counter to it and also occurs in a disc-bulge-halo system that doesn't form a bar. We provide a heuristic linear model that captures the essential dynamics of the instability.

\end{abstract}


%%%%%%%%%%%%%%%%%%%%%%%%%%%%%%%%%%%%%%%%%%%%%%%%%%

%%%%%%%%%%%%%%%%% BODY OF PAPER %%%%%%%%%%%%%%%%%%

\section{Introduction}

The most prominent manifestation of dark matter in disc galaxies such as the Milky Way is its contribution to the gravitational force necessary to explain the rotation curves of gas and stars in the disc. However, rotation curves constrain the spherically-averaged, radially-integrated dark matter density, but little else. There are, of course, non-axisymmetric phenomena, which may provide further constraints on the nature of dark haloes. In particular, between one and two thirds of all disc galaxies are barred \citep{eskridge2000, grosbel2004, sheth2008, erwin2018} and a similar fraction have prominent warps at large radii \citep{garciaruiz2002, reshetnikov2002, saha2009}. Theoretical arguments and N-body simulations have shown that properties of a halo, such as its shape, spin, and amount of substructure, can effect the formation and evolution of bars and warps. Yet despite over fifty years of research on this question, the dynamics of disc-halo systems is not fully understood.

Early investigations of the bar-halo connection, such as the seminal work of \citet{ostriker1973numerical}, showed that one could stabilize a disc against the bar instability by adding a rigid halo potential to the potential generated by the disc itself. It is now widely accepted that a proper treatment of bar formation in disc-halo systems requires that one model the halo as a system of "live" particles that can respond to changes in the gravitational field. For example the resonant coupling between disc stars and halo particles can provide an efficient means for transferring angular momentum from the inner disc to the outer disk and halo and thus help drive the bar instability \cite{lyndenbell1972, athanassoulareview,  weinberg1998, sellwood2016bar, petersen2016dark}. The expectation is then that bar formation in simulated galaxies will be less vigorous when a live halo is replaced by a rigid potential and more vigorous if we increase the number of halo particles that rotate in the same sense as the stars in the disc. These expectations have been borne out in a wide range of numerical experiments \citep{sellwood2016bar, polyachenko2016, saha2013, collier2019stellar,kataria2022effects}. By contrast, the instabilities that drive spiral structure seem to be hardly affected by whether the halo is live or static \citep{sellwood2021}. 

The velocity structure of the dark halo also appears to be important for the evolution of a bar after it forms. During the early stages of its evolution, a bar grows and its pattern speed decreases as it continues to shed angular momentum. However the growth of bars can be suppressed in spinning halos and in some cases, the halo can act as a source rather than sink of angular momentum for the disc \citep{long2014}. The upshot is that while the bar instability is more vigorous when the disc is embedded in a co-rotating halo than it would be in an isotropic halo, the subsequent growth in length and strength of the bar may be slower \citep{collier2019stellar}. 

The coupling between a bar and the dark halo can be viewed in terms of dynamical friction \citet{chandra1943,debattista1998dynamical}. As the bar sweeps through the halo, it forms a wake of halo particles that are scattered by its gravitational field. In fact, halo particles can be resonantly trapped by the bar thereby forming a shadow bar of dark matter \citep{hernquist1992, HB2005, athanassoula2007, petersen2016dark, collier2021coupling, chiba2022}. The presence of a shadow bar suppresses any further angular momentum transfer between the bar and the inner halo thereby curtailing growth of the bar.

The growth in the length and strength of a bar can be interrupted by buckling events, which cause a shortening and thickening of the bar \citep{combes1981, raha1991, debattista2004, martinez2004buckling, athanassoula2005}. As with bar formation itself, buckling appears to depend on the velocity structure of the halo. For example, \cite{li2022stellar} showed that the buckling instability in high spin prograde halos tends to be delayed. 

Bars are predominantly symmetric about the galactic mid plane and characterized by strong $m=2$ contributions to the surface density where $m$ is the azimuthal quantum number. By contrast, warps are predominantly anti-symmetric, $m=1$ features that arise from bending of the disc rather than a redistribution of mass. Warps can be generated by instabilities in the disc-halo system \cite{sparke1984galactic, sparke1988model, debattista1999warped} and tidal interactions with satellites \citep{weinberg2006magellanic}. In fact, weak warps can be generated spontaneously out of the noise in isolated galaxy \citet{chequers2017spontaneous, sellwood2022internally}.

As with bars, warps can couple to halos through dynamical friction \citet{bertin1980, nelson1995}. Typically dynamical friction will damp warps except in special circumstances, adding to the consensus that warps must be continually excited in the disc. One of these special cases is a retrograde halo, which may act to excite rather than damp warps since both halo and warp rotate in the same sense \citet{nelson1995}.

The aim of this paper is to explore the connection between halo rotation and the formation of bars and warps. To this end, we carry out four simulations each involving an identical disc embedded in either a static halo, a halo with isotropic orbits, or halos with prograde or retrograde rotation. In agreement with earlier studies, we find that bar formation occurs most rapidly in the prograde case, followed by the isotropic halo, the rotating halo, and static halo cases. Likewise, the shadow bar is strongest in the prograde case. We discuss in detail two types of buckling events. The first occurs soon after the bar forms and is asymmetric about the short axis of the bar with a 'tilde'-shaped vertical displacement profile. The second, which occurs $1-3\,{\rm Gyr}$ after the bar forms, is symmetric about the short axis of the bar and has a 'V'-shape displacement profile. Following \citet{li2022stellar}, we consider the circular patterns in the orbits of stars along the bar. We also study warps in our four simulations and confirm the theoretical conjecture from \citet{nelson1995} that warps are enhanced when the disc is embedded in a counter-rotating halo.

The most novel result from our work is the discovery of an instability that causes the center of mass of the disk and the cusp of the halo to spiral outward from the origin of the simulation. The displacement of disc and halo cusp from the origin grows exponentially with an e-folding time scale on the order $1-2\,{\rm Gyr}$ The effect is unrelated to bar formation, as demonstrated by the fact that it also occurs in a simulation with a bulge and a less massive and warmer disc that doesn't form a bar. We provide a heuristic linear model for the instability that qualitatively matches the dynamics. The effect is reminiscent of the sinking satellite problem in which a satellite galaxy loses orbital angular momentum and spirals inward due to dynamical friction from the halo particles (see \citet{chandra1943, lin1983, white1983, duncan1983, velazquez1999} for a sample of the many papers on this problem). In our case, the disc is swept outward by the rotating halo particles, which provide a reservoir of angular momentum. The rotating halo instability drives the system away from axisymmetry and therefore might play a role in generating lopsided features in spiral galaxies such as those described in \citet{jog2009} and references therein.

In Section \ref{sec:simulations}, we describe the initial conditions and N-body code for our simulations as well as some of the analysis tools used in subsequent sections. In Section \ref{sec:bars} we describe our results on bar formation, buckling events, and warps. In Section \ref{sec:rotating}, we present numerical evidence for the RHI along with the heuristic model. We summarize our results and present some concluding remarks in Section \ref{sec:conclusion}.

\section{Simulations}
\label{sec:simulations}

\subsection{initial conditions and N-body code}
\label{sec:nbodycode}

We carry out a sequence of N-body simulations that follow the evolution of a stellar disc embedded in a dark matter halo starting from equilibrium initial conditions. For the most part, we use the cylindrical coordinates $(R,\,\phi,\,z)$ though we sometimes switch to the spherical coordinates $(r,\,\theta,\,\phi)$ when discussing the halo. We assume that the disc has an exponential surface density profile $\Sigma = \Sigma_0\exp{(-R/R_d)}$ with a total mass of $M_d = 4.7\times 10^{10}\,M_\odot$, a radial scale length of $R_d = 2.8\,{\rm kpc}$, and a vertical scale height of $h_z = 180\,{\rm pc}$. The radial velocity dispersion profile is also assumed to be exponential with $\sigma_r = \sigma_{R0} \exp{(-R/2R_d)}$. We set $\sigma_{R0} = 100\,{\rm km\,s}^{-1}$, which gives a radial velocity dispersion of $24\,{\rm km\,s}^{-1}$ at $R=8\,{\rm kpc}$, the approximate position the Sun would have in this galaxy model. The density of the dark matter halo follows a Navarro-Frenk-White (NFW) profile \citep{navarro1996} with a scale length of $a_h = 13.6\,{\rm kpc}$ and a density at $r=a_h$ of $\rho_h = 2.72\times 10^6\,M_\odot\,{\rm kpc}^{-3}$, which translates to a halo velocity scale of $v_h = \left (16\pi G\rho a_h^2\right )^{1/2} \simeq 330\,{\rm km\,s}^{-1}$ as defined in equation 9 of \citet{widrow2008dynamical}.

All of our simulations are run with a live disc. In our fiducial simulation, the halo is also live and its velocity distribution is isotropic. We also follow the evolution of the disc in a static halo potential, in a halo that rotates in the same sense as the disc, and in one that rotates in the opposite sense of the disc. We refer to these simulations as IH, SH, PH, and RH for isotropic halo, static halo, prograde halo, and retrograde halo, respectively.

\begin{figure}\includegraphics[width=\columnwidth]{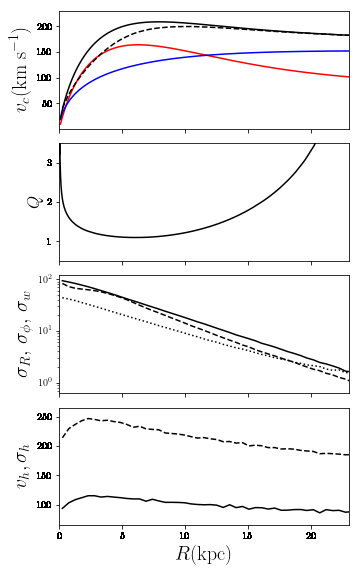}
    \caption{Circular speed and rotation curves, Toomre $Q$-parameter, and dispersion profiles for our models. The top panel shows $v_c(R)$ (solid black) as well as the separate contributions to $v_c$ from the disc (red) and halo (blue). The dashed black curve shows the rotation curve of the disc particles, which includes the the effect of asymmetric drift. The second panel shows the Toomre $Q$ parameter as a function of $R$. The third panel shows the velocity dispersion for disc particles in the radial, azimuthal, and vertical directions as solid, dashed, and dotted curves. The semi-log plot highlights the fact that the velocity dispersion profiles are approximately exponential in $R$. The bottom panel shows the total velocity dispersion of the halo (dashed curve) as well as the rotation curve for the prograde halo.}
    \label{fig:rotation}
\end{figure}

We have intentionally designed models that are susceptible to vigorous bar instabilities to allow us to study bar formation and buckling. The bars that form are, in general, stronger than the bar seen in the Milky Way. In the top panel of Fig.\ref{fig:rotation} we show the circular speed curve, $v_c(R)\equiv \sqrt{(R|F(R)|}$ where $F(R)$ is the radial component of the force as measured in the mid plane of the system. We also show the contributions to $v_c$ from the disc and halo. From the rotation curve decomposition, we can already anticipate that the system will form a strong bar since the disc dominates over the halo for $R<10\,{\rm kpc}$. To further emphasize this point we note that the ratio of the total circular speed to the disc's contribution is $X \equiv V_{\rm peak}/\sqrt{GM_d/R_d} \simeq 0.78$ whereas stability against the formation of a strong bar requires $X \ga 1.1$ \citep{efstathiou1982}. We also show the rotation curve, $\langle v_\phi\rangle(R)$ for disc particles. The difference between $v_c$ and $\langle v_\phi\rangle$ is a reflection of asymmetric drift, which is built into our initial conditions.

The susceptibility of the disc to instabilities is also evident in the second panel of Fig.\ref{fig:rotation} where we show the radial profile of the Toomre stability parameter $Q=\sigma_R\kappa/3.36 G\Sigma$ where $\kappa$ is the epicyclic frequency \citep{binney2008}. The epicyclic frequency descreases with $R$ as $R^{-\alpha}$ with $0<\alpha < 1$ depending on the shape of the rotation curve. The shape of the $Q$ profile is characteristic of disc-halo models (see, for example, figure 2 of \citet{sellwood1986} and figure 4 of \citet{kuijken1995nearly}). In our model, $\sigma_R^2\propto \Sigma\propto \exp{(-R/R_d)}$ and therefore $Q\propto R^{-\alpha}\exp{(R/2R_d)}$. The $Q$-parameter reaches a minimum of $Q_{\rm min}=1.1$ at $R\simeq 6\,{\rm kpc}$ indicated that this is a relatively cold disc, and hence one that will be susceptible to various instabilities.

The initial conditions are generated using the \textsc{GalactICS} code \citep{kuijken1995nearly, widrow2008dynamical, deg2019}, which is based on Jeans theorem where the distribution functions (DFs) for both the disc and halo are written in terms of elementary functions of the integrals of motions. The DF for the non-rotating halo is a function of the energy $E$ and therefore the orbits are isotropic. The DF for the disc is a function of two exact integrals of motion, $E$ and $L_z$, the angular momentum about the $z$-axis, and an approximate third integral of motion $E_z\equiv v_z^2/2 + \psi(R,z)-\psi(R,0)$ where $v_z$ is the $z$-component of the velocity $\psi$ is the gravitational potential. The quantity $E_z$ is quite well conserved along orbits that do not make large excursions from their guiding radius or the mid plane. This condition is satisfied for the vast majority of orbits in the cold discs considered here. Moreover, the system shows no evidence for disequilibrium or ringing even near the centre where the non-conservation of $E_z$ is most severe. This is in contrast with initialization methods based on the moment equations of the collisionless Boltzmann equation, where there are significant transients as the system relaxes from its initial state \citep{kazantzidis2004}.

The first step in the \textsc{GalactICS} code is to calculate the self-consistent DF-density-potential triad by solving Poisson's equation for the potential given the density and calculating the density from the DF. This is accomplished by expanding both the potential and density in Legendre polynomials and solving for the coefficients on a radial grid using an iterative scheme. We note that the isodensity contours of the halo are slightly flattened due to the contribution of the disc to the potential. The axis ratio of the halo in this work is $\simeq 0.994$; the difference between this and unity, though small, is enough to lead to a measurable precession in one of our experiments in Section \ref{sec:rotating}. In the SH simulation, we calculated the force due to the halo using the Legendre polynomial expansion and utility routines in the \textsc{GalactICS} code. The alternative is to simply freeze the halo particles, which avoids the necessity of integrating \textsc{GalactICS} routines into an N-body code \citep{polyachenko2016}.

An N-body realization for the disc and halo are obtained by sampling the DF as described in \citet{kuijken1995nearly}. For the PH and RH runs, we impart rotation to the halo by introducing a bias in the sign of the azimuthal component of the velocity, $v_\phi$, for halo particles. Since the halo DF depends on the square of $v_\phi$, we are free to choose any admixture of particles with positive and negative $v_\phi$. In this work, we choose the limiting case where all halo particles have positive $v_\phi$ in the PH case and negative $v_\phi$ in the RH case. This method for producing a rotating halo is rather artificial; more realistic models can be generated by allowing the DF to depend on the $L_z$ as well as $E$. Nevertheless, it leads to a halo with a reasonable amount of rotation. To see this, we compute the dimensionless spin parameter $\Lambda$ first discussed by \citet{peebles1969, peebles1971}. Here, we use the form from \citet{bullock2001}:
\begin{equation}
    \Lambda = \frac{|\langle L_z\rangle|}{\sqrt{2GM_{200}R_{200}}},
\end{equation}
where $\langle L_z\rangle$ is the mean angular momentum of the halo particles, $R_{200}$ is the virial radius of the halo, defined to be the radius in which the mean density interior is 200 times the critical density, and $M_{200}$ is the virial mass. For our halo, $R_{200} \simeq 145\,{\rm kpc}$ and $M_{200} \simeq 3.3\times 10^{11}\,M_\odot$. With these values, the PH and RH models have a spin parameter of $\Lambda \simeq 0.08$. The halos in cosmological simulations roughly follow a log-normal distribution with a mean of $0.04$ and standard deviation in $\ln{\Lambda}$ of 0.5 \citep{bullock2001}. Our PH and RH halos therefore have high spins but within 1.5 sigma of the mean. 

We stress that the initial conditions as generated by the \textsc{galactICS} code are in dynamical equilibrium, that is, represent a solution to the time-independent collisionless Boltzmann and Poisson equations, apart from the approximation that $E_z$ is an integral of motion and the Poisson noise of the N-body sample. For the relatively cold discs considered in this paper, $E_z$ is very nearly conserved \citep{kuijken1995nearly, widrow2008dynamical, deg2019}.

The third panel of Fig.\ref{fig:rotation} shows the radial profiles for the velocity dispersion in the radial, azimuthal, and vertical directions as calculated from the N-body realization of the disc. By design, the radial dispersion profile is an exponential with scale length $R_d/2$ (see discussion above). The azimuthal dispersion is given by $\sigma_\phi = \left (\kappa/2\Omega\right )\sigma_R $ where $\Omega$ is the angular frequency (see, for example, \citet{binney2008}). Finally, the vertical velocity dispersion is determined by solving the collisionless Boltzmann and Poisson equations under the assumption that the vertical structure is approximately isothermal and the thickness of the disc is approximately constant in radius.

The bottom panel of Fig.\,\ref{fig:rotation} shows the total velocity dispersion profile for the halo as derived from the N-body realization of the equilibrium DF. As expected, the peak in the 3D velocity dispersion is roughly a factor of $\sqrt{3/2}\simeq 1.22$ times higher than the peak in the rotation curve. We also show the rotation curve for the halo in the PH case. The rotation curve is about a factor of two lower than the rotation curve for the disc stars reflecting the fact that even for our limiting PH and RH cases, the main support for the halo comes from random motions of its constituent particles.

The simulations are performed using an N-body code from \citet{stiff2003}, which is based on the octree algorithm described in \citet{dehnen2000}. The systems were evolved for $10\,{\rm Gyr}$ using a timestep of $\simeq 1\,{\rm Myr}$ and a softening length of $50\,{\rm pc}$. The main suite of simulations were run with 2M particles in both disc and halo components. In general, using more particles would delay the onset of the bar instability but not change the rate at which the bar instability grows (see, for example \citep{dubinski2009}). As a test of momentum conservation, we followed the velocity of the center of mass and found that it was $<10^{-3}\,{\rm km\,s}^{-1}$ with no systematic drift. Likewise, the total angular momentum was conserved to better than a part in $10^3$.

\subsection{analysis tools}

Throughout this work, we use Fourier analysis to describe the angular and time dependence of various quantities such as the surface density and mid plane displacement. The methods are described in \citet{sellwood1985}, \citet{athanassoula2002}, and \citep{chequers2017spontaneous}. We begin by dividing the disc into cylindrical bins labelled by $\alpha$ with mean radii $\{R_\alpha\}$ and areas $\{S_\alpha\}$. The surface density is then estimated to be
\begin{equation}
    \Sigma(R_\alpha,\,\phi,\,t) = 2\pi S_\alpha^{-1}
    \sum_{j\in\alpha} \mu_j \delta(\phi - \phi_j(t))
\end{equation}
where the sum is over all particles in the $\alpha$'th bin. This can be expressed as a Fourier series:
\begin{equation}
    \Sigma(R_\alpha,\,\phi,\,t) = \sum_{m=0}^\infty \Sigma_m(R_\alpha,\,t)
    e^{-im\phi}
    \label{eq:sigmaRphit}
\end{equation}
where it is understood that $\Sigma$ is obtained by taking the real part of the right hand side of this expression. The Fourier coefficients are given by
\begin{equation}
    \Sigma_m(R_\alpha,t) = S_\alpha^{-1} \sum_{j\in \alpha} \mu_j e^{im\phi}.
\end{equation}
The Fourier series allows us to construct an approximation to the surface density that is continuous in $\phi$. In what follows, we use terms up to $m=5$ and write
\begin{equation}
    \bar{\Sigma}(R_\alpha,\phi,t) = \sum_{m=0}^5
    \Sigma_m(R_\alpha,\,t) e^{-im\phi}.
\end{equation}
We can construct similar quantities to describe the displacement of the disc from its mid plane and the thickness of the disc. For example, the vertical displacement across the disc is given by
\begin{equation}
    Z(R,\phi,t) = \bar{\Sigma}(R,\phi,t)^{-1}
\sum_m Z_m(R_\alpha,\,t) e^{-im\phi}    
\end{equation}
where
\begin{equation}
    Z_m(R_\alpha,\,t) = S_\alpha^{-1} \sum_j \mu_j z_j e^{im\phi_j}
\end{equation}

Global properties of the disc are described by considering bins that encompass large regions of the disc. For example, we define the dimensionless bar strength parameter as 
\begin{equation}
    A_2 = 
    M_{7}^{-1} \left |\sum_{R<7\,{\rm kpc}} \mu_j e^{im\phi_j}\right |
\end{equation}
where $M_{7}$ is the mass of the disc within $7\,{\rm kpc}$. As discussed below, we choose $7\,{\rm kpc}$ to divide inner and outer regions of the disc since this radius corresponds to the corotation radius of the bar when it first forms. We similarly define the global Fourier mode with quantum number $m$ associated with vertical displacements as
\begin{equation}
    Z_m = 
    M_{7}^{-1} \left |\sum_{R<7\,{\rm kpc}} \mu_j z_j e^{im\phi_j}\right |
\end{equation}
Finally, the root-mean-square thickness of the disc is defined as 
\begin{equation}
    Z_{\rm RMS} =  \sqrt{M_{7}^{-1} \sum_{R<7\,{\rm kpc}} m_j z_j^2}
\end{equation}

\section{Bars and Warps}
\label{sec:bars}

\subsection{formation and growth of the bar}

\begin{figure}
	\includegraphics[width=\columnwidth]{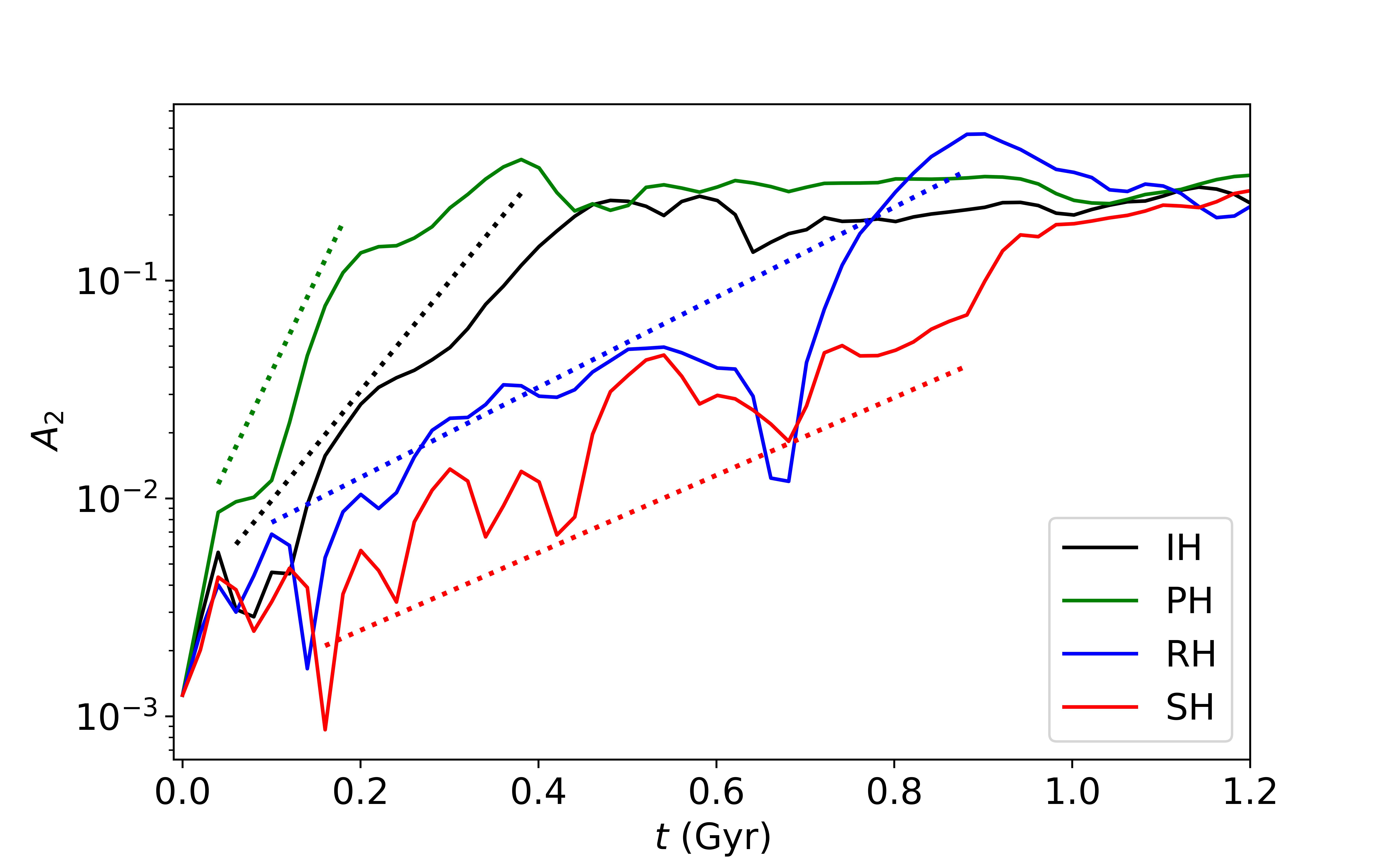}
    \caption{Bar strength parameter $A_2$ as a function of time for the first $1.2\,{\rm Gyr}$ of the simulation. The solid lines show $A_2$ for the IH (black), PH (green), RH (blue) and SH (red) simulations. The dotted lines indicate exponential growth $A_2\propto \exp{(\Gamma_2 t)}$ at these early times. The rates are $\Gamma_2\simeq 5.0, 8.6, 2.1, 1.8\,{\rm Gyr}^{-1}$ for the IH, PH, RH, and SH cases, respectively.}
    \label{fig:A2}
\end{figure}

We begin by examining bar formation in our four simulations. In Fig.\,\ref{fig:A2} we show the evolution of $A_2$ for $t \le 1.2\,{\rm Gyr}$. In all cases, $A_2$ grows roughly as an exponential function of time, saturating at values between $0.25$ and $0.4$, thereby signaling the formation of a strong bar. We estimate the growth rates to be $8.6, 5.0, 2.1, 1.8\,{\rm Gyr}^{-1}$ for the PH, IH, RH, and SH simulations, respectively. As expected, the bar instability is most vigorous when the disc is embedded in a co-rotating halo and least vigorous when the halo is replaced by a rigid potential.

A more complete picture of bar formation is provided in Fig. \ref{fig:magnitudes}. In the top panel we show $A_2$ over $10\,{\rm Gyr}$. Consider first, the IH case. The initial sharp rise in $A_2$ already seen in Fig.\,\ref{fig:A2} is followed by a period extending from $0.5 {\rm Gyr}$ to $3.7\,{\rm Gyr}$ that is characterized by rapid oscillations superimposed onto steady growth.  At $3.7\,{\rm Gyr}$, $A_2$ drops by about 25\% after which there is a second period of steady growth, reaching $0.35$ at $10\,{\rm Gyr}$.

Further insight into the sequence of events for the IH simulation is gained in the second and third panels where we show the $Z_2$ and $Z_{\rm RMS}$. At $t\simeq 1\,{\rm Gyr}$, $Z_2$ reaches a peak of $10\,{\rm pc}$ indicating that the nascent bar is bending in and out of the disc plane, albeit by a relatively small amount. There is a more prominent peak with $Z_2\simeq 60\,{\rm pc}$ at $t\simeq 3.7\,{\rm Gyr}$ after which $Z_2$ decays exponentially. The thickness of the disc increases throughout the simulation with two periods where the rate of increase is especially rapid, one at the start of the simulation when the bar is forming and one at $3.7\,{\rm Gyr}$. As discussed below, we identify the peaks in $Z_2$ and associated decreases in $A_2$ and rapid increases in $Z_{\rm RMS}$ as buckling events.

The evolution of $A_2$, $Z_2$, and $Z_{\rm RMS}$ is similar in the PH and RH halos though the timing of the various events is different. For example, the initial buckling event occurs later in the RH case and earlier in the PH case as compared with the IH case. Conversely, the second buckling event occurs later in the PH case and earlier in the RH case. Finally, the disc in the SH simulation has the weakest bar and least amount of bending relative to the mid plane or vertical heating.

One should keep in mind that $A_2$ is not a perfect diagnostic for bar formation since it also has contributions from two-armed spirals. Indeed, the short term oscillations seen in all of the simulations are likely the result of beats between two features with different pattern speeds such as spiral arms and the bar \citep{villa2009dark,sellwood2016bar}.

\subsection{angular momentum transport}

\begin{figure}
	\includegraphics[width=\columnwidth]{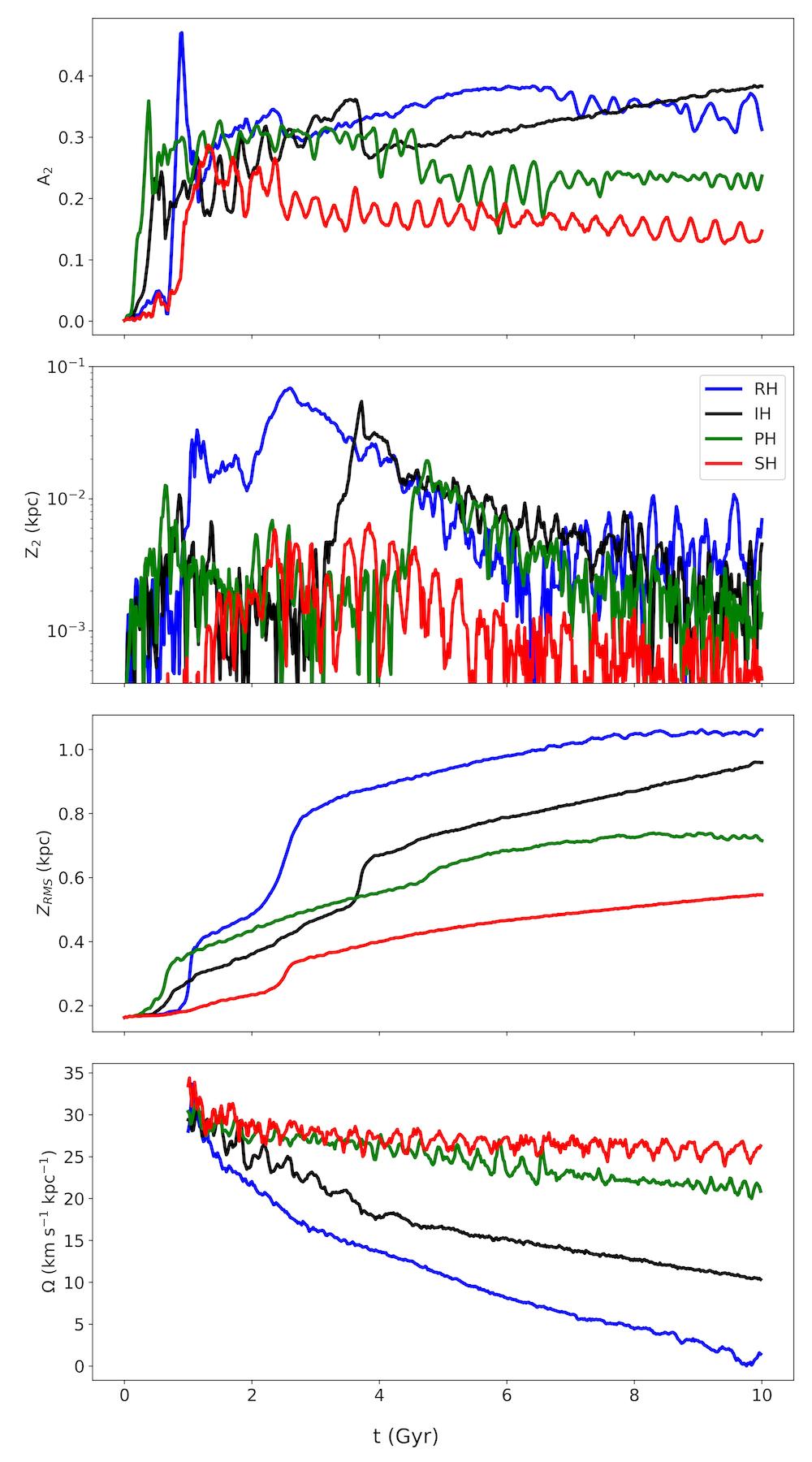}
    \caption{Global properties of the disc as a function of time. The line colors are the same as in Fig.\,\ref{fig:A2}. From top to bottom, we show the bar strength $A_2$, the mean $m=2$ mode of vertical displacement $Z_{2}$, the root mean square thickness of the disc, $Z_{\rm RMS}$, and the bar pattern speed, $\Omega$. As discussed in the text, all quantities are calculated using particles within $7\,{\rm kpc}$ of the centre of the disc.}
    \label{fig:magnitudes}
\end{figure}

The bottom panel of the Fig. \ref{fig:magnitudes} shows the bar pattern speed $\Omega$ for $t\ge {\rm Gyr}$. In all four simulations, the bar forms with $\Omega \simeq 35\,{\rm km\,s}^{-1}\,{\rm kpc}^{-1}$, which corresponds to a corotation radius of $\sim 7\,{\rm kpc}$. The pattern speed then decreases with a spin-down rate that is most rapid in the RH case and least rapid in the PH and SH cases. These results are in qualitative agreement with those of \citet{collier2021coupling}. Conversely, \citet{saha2013} found that the spin down rate of the bar was strongest when the disc was embedded in a high-spin prograde halo and weakest in their retrograde and non-rotating runs. It is worth noting that the disc in the \citet{saha2013} simulations were relatively light compared to ours ($X=V_{\rm peak}/\sqrt{GM_d/R_d}\simeq 2$ for their system as compared with $0.78$ for ours) and warm ($Q\simeq 2$ throughout much of their disc as compared with $1.1$ in ours). On the other hand, \citet{collier2021coupling} used a slightly heavier disc than we did and a radial velocity dispersion profile very similar to ours. Since the coupling between the bar and the halo depends both on the mass of the bar and the relative speed between the bar and the halo particles, it is perhaps not surprising that the spin-down rates of the bars in the \citet{saha2013} simulations were different from spin-down rates in our simulations and those of \citet{collier2019stellar}.

\begin{figure}
	\includegraphics[width=\columnwidth]{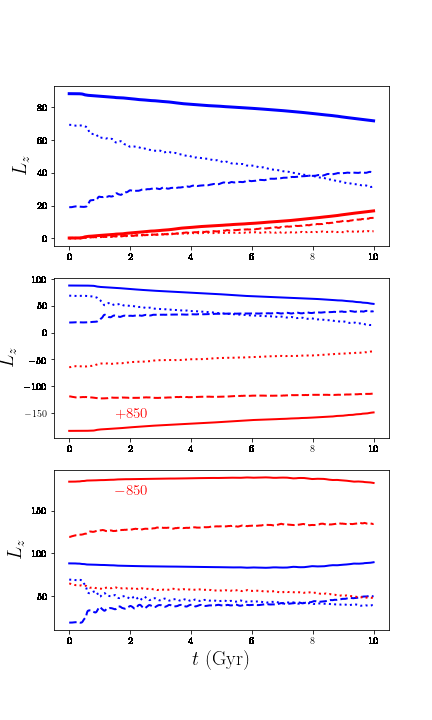}
    \caption{Angular momentum about the $z$-axis $L_z$ in different subcomponents of the galaxy as a function of time. Top, middle, and bottom panels are for the IH, RH, and PH cases, respectively. Blue curves are for the disk; red curves are for the halo. The dotted curves are for particles within $7\,{\rm kpc}$; Dashed curves are for the particles outside this radius; solid curves are for the entire system. The units for $L_z$ are $10^{11}\,M_\odot {\rm kpc} \,{\rm km\,s}^{-1}$. The numbers next to the solid and dashed lines for the halo give the offsets that allow the various curves to be shown on the same figure.}
    \label{fig:jz}
\end{figure}

As discussed in the introduction, angular momentum transport plays a central role in the formation and evolution of bars. In Fig.~\ref{fig:jz} we show the change in total angular momentum of the inner and outer disc and halo as defined by $R = 7\,{\rm kpc}$. In each of the simulations, the inner disc loses angular momentum and the outer disc gains angular momentum. This angular momentum transport is likely due to the torque on the bar from trailing spiral arms in the outer disc. In the IH case, both the inner and outer halo gain angular momentum. By contrast, in the PH case, the inner halo actually loses angular momentum. As discussed below, this is the case where a strong shadow bar forms in the dark halo.

Our findings of angular momentum transfer are consistent with the bar pattern speed panel of Fig.~\ref{fig:magnitudes}. The SH bar experiences the least amount of spin down since there is no dynamical friction acting on the bar due to the halo. Put another way, the halo is not available to act as an angular momentum sink for the bar. Dynamical friction is strongest in the RH case. Conversely, in the PH case, a shadow bar forms in the inner halo, which implies that the inner halo is losing rather than gaining angular momentum.

\subsection{dark bar}

\begin{figure}
	\includegraphics[width=\columnwidth]{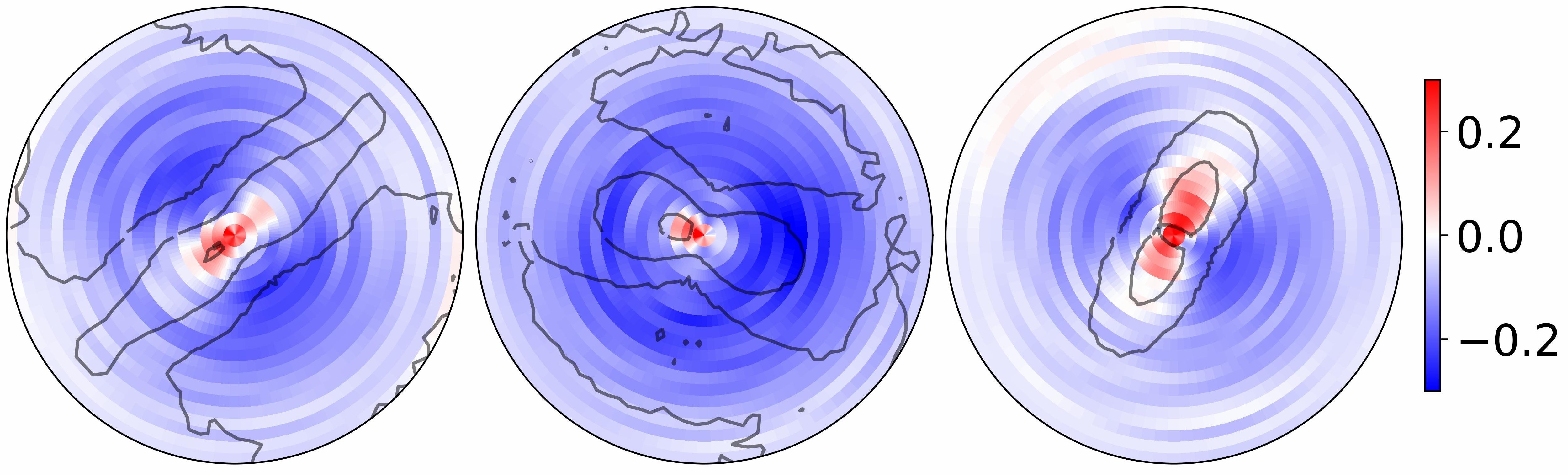}
    \caption{Face on views of the halo density in the $z=0$ plane at $8\,{\rm Gyr}$ normalized by the initial halo density on a log base ten scale. We show, from left to right, the results for the IH, RH, and PH simulations. Black contours provide a coarse picture of the surface density in the disc and allow one to visualize the position of the bar.}
    \label{fig:darkbar}
\end{figure}

It is now well-established that the formation of a bar in a stellar disc will change the structure of the halo. In particular, the inner cusp of the halo becomes elongated leading to the concept of a dark or shadow bar (see, for example, \cite{hernquist1992, HB2005, athanassoula2007, petersen2016dark, collier2021coupling}).  Here we explore the distortions of the inner halo in our rotating halo simulations. In Fig. \ref{fig:darkbar}, we show face on views of the halo density in the $z=0$ plane. The density in the mid plane is calculated by first writing the halo density in terms of spherical harmonics
\begin{equation}
    \rho_h(r,\theta,\phi) = \sum_{l,m}\rho_{l,m}(r) Y_{lm}(\phi,\theta).
\end{equation}
The radial coefficients $\rho_{l,m}$ are determined by summing over particles weighted by the complex conjugate of the $Y_{lm}'s$ \citet{binney2008}. The desired density at $z=0$ is then given by
\begin{equation}
    \rho_h(R,\phi,z=0) = \sum_{l,m} \rho_{l,m}(R) Y_{lm}(\phi,\theta=\pi/2).
\end{equation}
where we sum terms up to $l = 4$. We find that a bar-like structure, which is aligned with the stellar bar, forms in all three live halo simulations, as expected from previous studies of bar formation in disc-halo systems. The most prominent dark bar forms in the PH case as one might expect from the fact that this case has the strongest coupling between halo and disc particles. The dark bar in the IH is significantly shorter and the dark bar in the RH case is asymmetric about the center of the disc and the weakest of the three cases.

\subsection{Buckling Events}

As discussed above, the sharp peaks in $Z_2$ and the peaks in the growth rate of $Z_{RMS}$ seen in Fig. \ref{fig:magnitudes} signal events when the bar bends in and out of the mid plane of the disc. In each of our live halo simulations, there is a buckling event that occurs right as the bar forms and one that occurs after the bar has had a chance to settle down and grow in length and strength. We refer to the former as early buckling events and label them with the subscript $B1$. Likewise, we refer to the latter as late buckling events and label them by the subscript $B2$. We identify the times of these events using $Z_{\rm RMS}$, which is not as noisy as $Z_2$. The events, denoted $\rm IH_{B1}$, $\rm IH_{B2}$, $\rm PH_{B1}$, $\rm PH_{B2}$, $\rm RH_{B1}$ and $\rm RH_{B2}$, occur at 680 and 3700, 600 and 4800, 1000 and 2500 $Myr$, respectively. In the SH simulation there is no discernible peak in $Z_2$, though there is a small but sharp jump at 2500 $Myr$ in $Z_{RMS}$. We refer to this event as $\rm SH_{B1}$. As we'll see below, the buckling event in the SH simulation has the same morphology as the early-type buckling events in our live halo simulations. 

\begin{figure*}
    \centering
	\includegraphics[width=\textwidth]{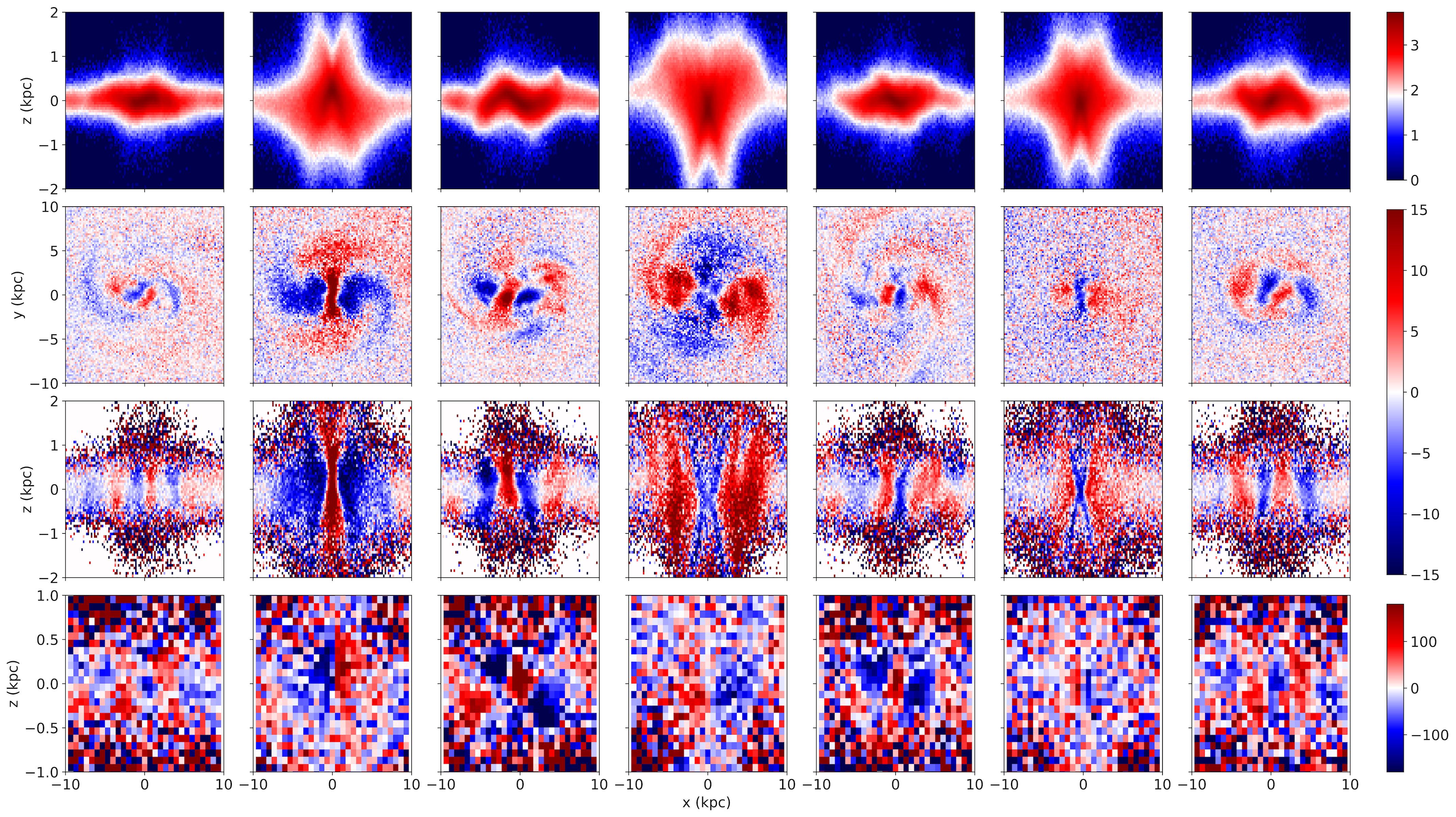}
    \caption{Moments of the disc DF in the $xz$ and $xy$ planes. From left to right each column shows results for the $\rm IH_{B1}$, $\rm IH_{B2}$, $\rm RH_{B1}$, $\rm RH_{B2}$, $\rm PH_{B1}$, $\rm PH_{B2}$ and $\rm SH_{B1}$ buckling events. The top row shows the surface density on a log base ten scale for a slab $2\,{\rm kpc}$ in width through the long axis of the bar and perpendicular to the disc plane. The second row shows the mean vertical velocity $\langle v_z\rangle$ in the disc ($xy$) plane. The third and fourth rows show $\langle v_z\rangle$ and the vorticity $\omega_y$ in the same plane used for the surface density in the top row. The color bar for $\langle v_z\rangle$ are in ${\rm km}\,{\rm s}^{-1}$ and the color bar for the vorticity is in ${\rm km}\,{\rm s}^{-1}\,{\rm kpc}^{-1}$.}
    \label{fig:bucklevz}
\end{figure*}

Following \citet{li2022stellar, li2023}, we map, in Fig. \ref{fig:bucklevz}, various moments of the disc DF at the seven buckling events identified above. For each snapshot, we orient the $x$ axis to coincide with the long axis of the bar. The top row shows the surface density across the $xz$-plane as calculated for a slab $|y|<2\,{\rm kpc}$. The distinction between the early buckling events and the later buckling events is readily apparent. The B1 events (first, third, fifth, and seventh columns) have a clear "tilde"-shape that is anti-symmetric about short axis of the bar, that is, $x=0$ \citep{patsis2002orbital, skokos2002orbital}. By contrast, with the late buckling events (second, fourth, and sixth columns) the surface density has a "V"-shape that is symmetric about the short axis of the bar. The dichotomy between the two types of events is also evident in the second and third rows, where we plot the mean vertical velocity $\langle v_z\rangle$ in the $xy$ and $xz$ planes. For the early events, the pattern is somewhat disorganized but predominantly antisymmetric about the short axis of the bar. With the late events, the pattern is more organized and symmetric about the short axis of the bar. The strongest late buckling event is the RH case and weakest in the PH case. Furthermore, the vertical velocity map during the $\rm RH_{B2}$ event has more substructure than in the $\rm IH_{B2}$ and $\rm PH_{B2}$ cases. As noted above, the bar in the RH case formed last but experienced its main, symmetric buckled event first.

\citet{li2022stellar} emphasized the existence of vortical flows along the bar in what we are referring to as late buckling events. In the last row in Fig.\ref{fig:bucklevz}, we show the vorticity in the $y$-direction, $\omega_y = \partial v_z/\partial x - \partial v_x/\partial z$. As in \citet{li2022stellar} we find an anti-symmetric pattern in the vorticity for our late buckling events though the pattern is realitively weak in the PH case. In other words, the bar is rotating about the $y$-axis in opposite directions on either side of the bar, in agreement with the $V$-shape pattern in density seen in the top row and the vertical motions seen in the second and third rows. The magnitude of the vorticity is of order $100 \, {\rm km\,s^{-1}}\,{\rm kpc}^{-1}$, which is quite large. The vertical velocities along the bar are of order $5-10\,{\rm km}\,{\rm s}^{-1}$ and change sign over a scale length of about $5\,{\rm kpc}$, implying a vorticity of $2-4\,{\rm km}\,{\rm s}^{-1}$. The main contribution to the vorticity actually comes from the vertical gradient in the radial ($x$) component of the velocity field.

The vorticity pattern for our early buckling events is more difficult to discern but does show the expected pattern based on surface density and vertical velocity maps. The inner bar rotates in one sense while the outer bar on either side of the center of the galaxy rotates in the opposite sense. 

Our late buckling events have an $m = 2$ structure and have been associated with orbits being trapped in the 2:1 vertical resonance \citep{lokas2019anatomy}. However, other resonances are possible such as the 3:1 vertical resonance, which can look like a tilde edge-on \citep{patsis2002orbital, skokos2002orbital}. These resonances may help explain the early type buckling events seen in Fig.~\ref{fig:bucklevz}. If a substantial number of orbits are trapped in a resonance then that could trigger to a cohesive response thereby driving the buckling instability \citep{li2023}. 

The standard explanation for buckling is that it is the result of a firehose or bending instability. Bending instabilities in razor thin infinite slabs are discussed in \citet{toomre1966kelvin} and \citet{kulsrud1971} while a more detailed study that includes the finite thickness of the slab is found in \citet{araki1985}. These works conclude that a slab is unstable to bending instabilities provided $\sigma_z/\sigma_R<0.3$ where $\sigma_z$ and $\sigma_R$ are the vertical and in plane components of the velocity dispersion. \citet{merritt1994} argue that for a disc rather than infinite slab, the less stringent condition of $\sigma_z/\sigma_R<0.6$ for the bending instability is more apt. 

\begin{figure}
    \centering
	\includegraphics[width=0.7\columnwidth]{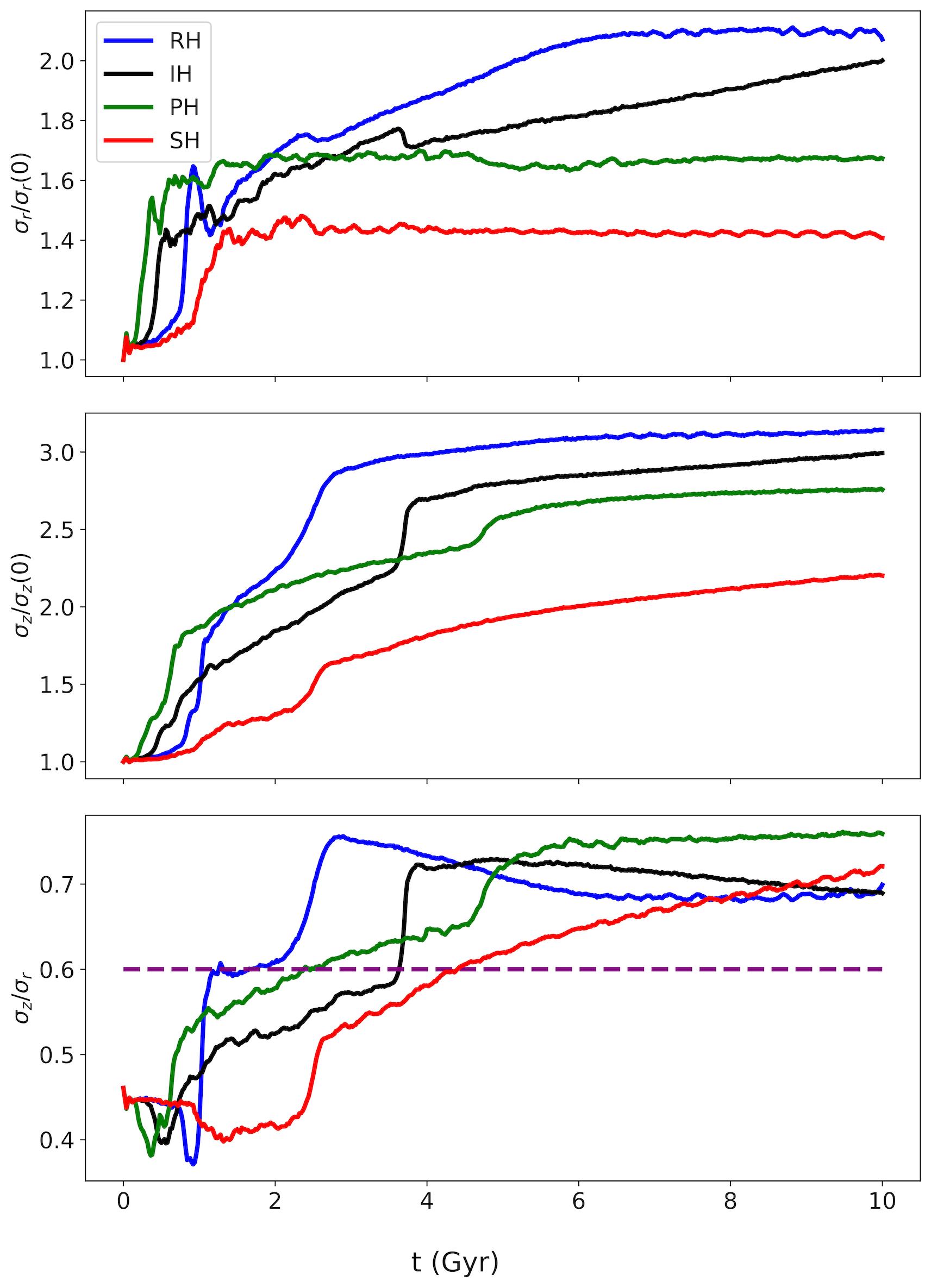}
    \caption[Velocity dispersion ]{From top to bottom: the radial velocity dispersion normalized by the radial dispersion at $t =0$, the vertical velocity dispersion normalized by the vertical dispersion at $t =0$, the ratio of vertical velocity dsipersion to radial velocity dispersion. Disc particles within two radial scale lengths were used. }
    \label{fig:dispersions}
\end{figure}

To test these results, we examine the time-dependence of the radial and vertical velocity dispersion for our four simulations. In the top panel of Fig.~\ref{fig:dispersions}, we plot the mean radial velocity dispersion for the entire disc $\sigma_R$ normalized by its initial value $\sigma_{R0}$. At early times, $\sigma_R/\sigma_{R0}$ rapidly increases by $40-60\%$. As one would expect, the timing of this increase mirrors the increase in $A_2$. In the SH and PH cases, $\sigma_R$ is approximately constant thereafter. On the other hand $\sigma_R$ continues to increase throughout the simulations in the RH and IH cases, except for small drops near the buckling events.

The middle panel of Fig.~\ref{fig:dispersions} shows the vertical velocity dispersion $\sigma_z$ again normalized by its initial value. $\sigma_z$ increase monotonically in all simulations with periods or particular rapid rise that coincide with buckling events. The bottom panel of Fig.~\ref{fig:dispersions} shows the ratio $\sigma_z/\sigma_R$. We also indicate the condition from \citet{merritt1994} for the buckling instability. In all cases, $\sigma_z/\sigma_R$ drops from its initial value of $0.45$ when the bar forms and then rises rapidly during the early type buckling events. The subsequent evolution of $\sigma_z/\sigma_R$ seems to depend sensitively on halo rotation. For example, in the IH and PH cases, $\sigma_z/\sigma_R$ increases gradually between buckling event with the late event occurring when $\sigma_z/\sigma_R$ has reached $0.55$ and $0.65$, respectively. In the RH case, $\sigma_z/\sigma_R$ is approximately constant at a value of $0.6$ for $1\,{\rm Gyr}$ with both $\sigma_z$ and $\sigma_R$ increasing at about the same rate. We conclude that bars can exist below the stability line for $\sigma_z/\sigma_R$ for $1-3\,{\rm Gyr}$. Moreover, buckling can occur in bars with $\sigma_z/\sigma_R\gtrsim 0.6$. Both of these findings point to the fact that buckling events cannot be fully explained by kinematic properties of the bar alone. 

\subsection{Warps}

In this section, we describe the warps that arise in our suite of simulations. In general, warps can form by various mechanisms such as tidal interactions with a passing satellite galaxy or the triaxial halo that surrounds the disc \citet{kahn1959, hunter1969dynamics, sparke1984galactic, sparke1988model, binney1992, debattista1999warped, shen2006galactic, revaz2004, chequers2018}. Though these mechanisms don't apply to our simulations, theoretical arguments and previous simulations suggest that warps can arise in isolated, nearly axisymmetric disc-halo systems. For example, \citet{bertin1980} argued that warps can be excited through a disc-halo bending instability that involves the transfer of angular momentum from disc to halo. A more careful treatment by \citet{nelson1995} found that halos can either damp or excite warps depending on the orbits of the halo particles. On the other hand, simulations have shown that that relatively weak warps could be generated spontaneously from axisymmetric initial conditions for disc-halo system \citet{chequers2017spontaneous, sellwood2022internally}. The explanation for these warps appears to be the misalignment between the inner and outer discs, which results in outwardly travelling, leading, retrograde bending waves \citet{shen2006galactic, sellwood2022internally}.

\begin{figure}
    \centering
	\includegraphics[width=0.7\columnwidth]
 %{figures/bendingfo.jpg}
 {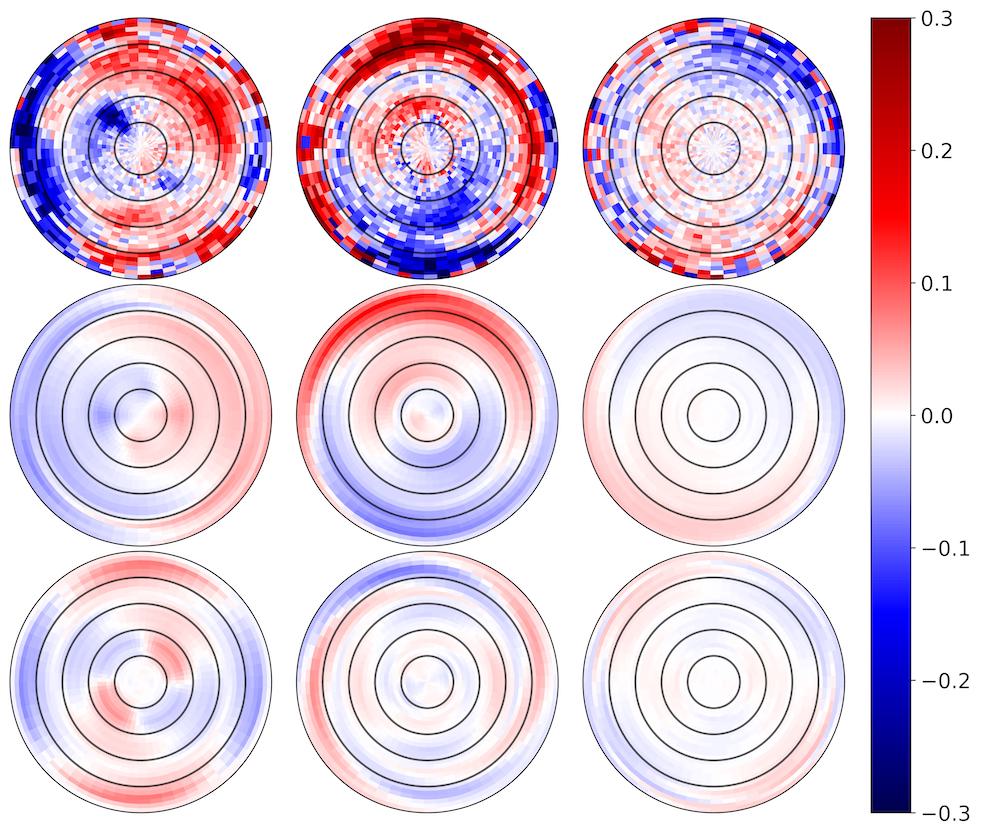}
    \caption[Face-on plots of mean z]{Mean vertical displacement in the disc plane at 10 Gyr for each of our live simulations. The disk rotates in the counterclockwise direction. From left to right: The IH, RH, and PH models. Shown are mean $z$ (top row) and $m=1$ and $m=2$ Fourier models. The color bar units are in ${\rm kpc}$. Black circles are drawn every 5 kpc.}
    \label{fig:bending}
\end{figure}

\begin{figure}
    \centering
	\includegraphics[width=0.7\columnwidth]{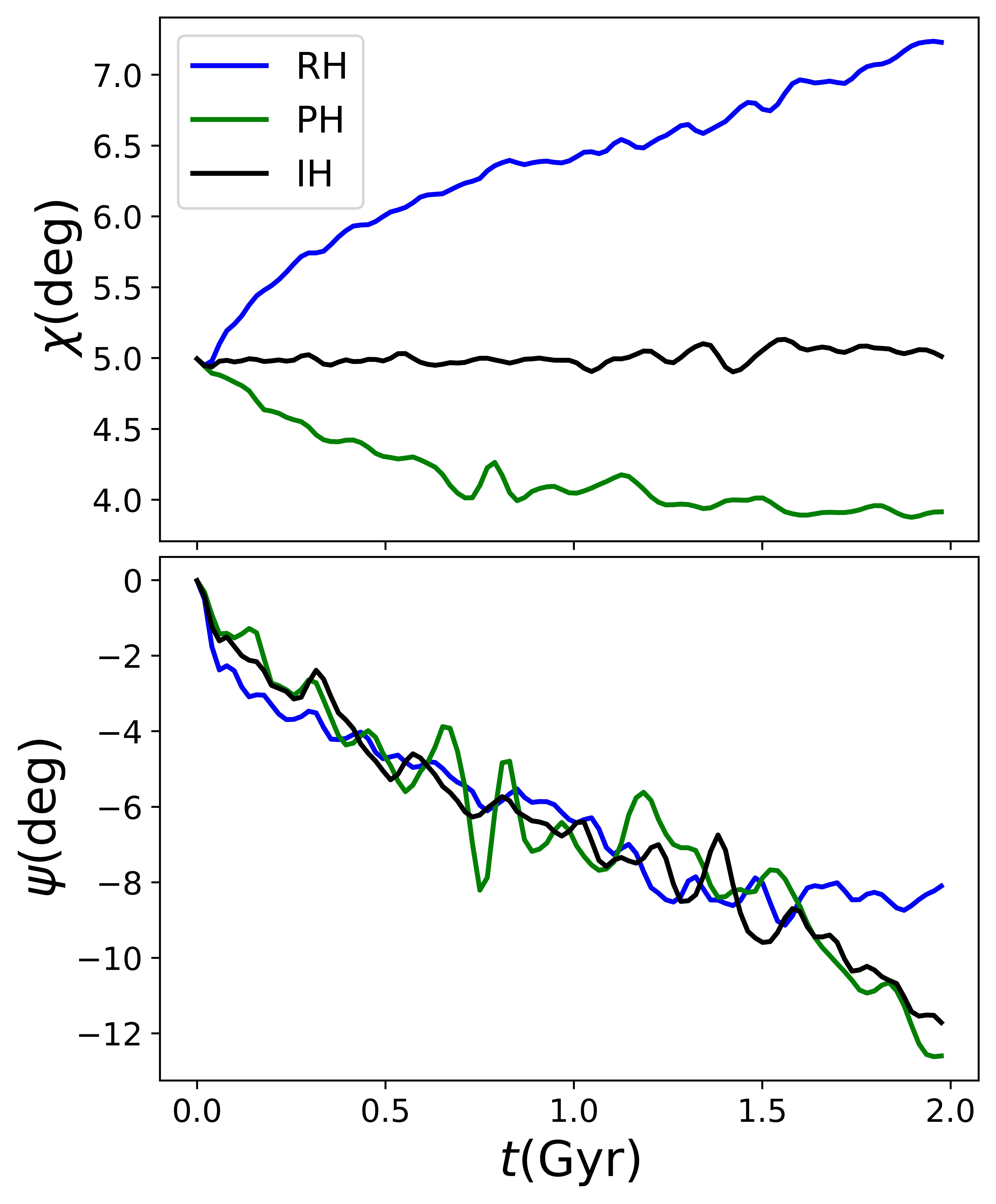}
    \caption[Inclination and precession angles]{Inclination ($\chi$) and precession ($\psi$) angles of the disc for the simulations where the initial inclination angle is $5^\circ$. A negative $\psi$ indicates that the disc is precessing opposite to the direction of Galactic rotation. As before, line colors are black, green, and blue for IH, PH, and RH, respectively.}
    \label{fig:tilt}
\end{figure}

As with bars, warps are affected by dynamical friction as they rotate through the dark halo though there has been some debate over whether a live halo will damp or excite and sustain warps \citet{bertin1980, nelson1995, debattista1999warped}. The conclusion in \citet{nelson1995} is that haloes damp warps unless the halo is counter-rotating. A simple heuristic argument is that dynamical friction tries to bring the angular momentum vectors of the halo and disc into alignment. Thus warps will be strongly damped in co-rotating haloes but enhanced in counter-rotating ones. These conclusions are confirmed in Fig. \ref{fig:bending} where we show the mean vertical displacement along with $m=1,2$ Fourier modes as a across the disc plane for our three live halo simulations. The RH case shows a strong, organized warp whose dominant structure is that of an $m=1$ leading bend in the disc with a pitch angle of about $0.07\,{\rm rad}$ or about $4^\circ$. There is also a significant contribution from a leading $m=2$ bend. The bending in the IH is more disorganized and clearly has large contributions from $m>2$ terms. The PH case shows the weakest bends.

To further explore the effect of halo rotation on warps, we rerun our three live halo simulations  with the disc initially tilted by $5^\circ$ with respect to the midplane of the halo. Fig.~\ref{fig:tilt} shows the evolution of the inclination angle ($\chi$) and precession angle ($\psi$) over 2 Gyr. The inclination angle of the disc increases in the RH case, decreases in the PH case, and stays roughly constant in the IH case. Interestingly enough, the time-dependence of the procession angle is very similar in the three cases with a rate of precession $\Omega_p\simeq 6^\circ \,{\rm Gyr}^{-1}\simeq 0.1\,{\rm rad}\,{\rm Gyr}^{-1}$. The precession arises because the symmetry axis of the disc is misaligned with that of the slightly-flattened halo (see Section \ref{sec:nbodycode}). From dimensional analysis, we have that the torque of the halo on the disc is $\tau \sim I_d\Omega_p^2$ where $I_d=6M_dR_d^2$ is the moment of inertia for an exponential disc. On the other hand, the radial force of the halo on the  disc is $F\simeq M_d v_h^2 R^{-1}$ where $v_h$ is the contribution of the halo to the circular speed at radius $R$. Thus, the dimensionless ratio $6R_d^2\Omega_p^2/v_h^2\simeq 3\times 10^{-5}$. Recall that the axis ratio of the halo was $\simeq 0.995$, which corresponds to a flattening of $0.005$. This is roughly the square root of the ratio derived above, as expected.

\section{Rotating Halo Instability}
\label{sec:rotating}

In general, the dark matter haloes found in cosmological N-body simulations are tumbling, triaxial systems with axis ratios in the range $0.5-0.9$ (see, for example, \citet{maccio2008}). And while most of the halo particles are distributed in a smooth background, about $5-10 \%$ of the total halo mass resides in subhaloes \citep{gao2004}. We therefore expect that the gravitational force on a disc due to the halo in which it resides will have non-axisymmetric, time-dependent components that will cause the disc to wander relative to the center of mass of the system as a whole. Evidence for this effect has been seen in simulations \citep{kuhlen2013} as well as observations of merging galaxies \citep{jog2006}. A particularly striking example is the reflex motion of the Milky Way's disc due to the Large Magellanic Cloud, which can reach tens of kilometers per second \citep{petersen2021}.

In our simulations, the dominant departures from axisymmetry are the stellar and dark bars. We therefore expect that any motion of the disc away from the centre of mass of the system should be small since these are mainly even-$m$ perturbations to the mass distribution and potential. Of course, Poisson noise in the simulation will cause the disc centre to wobble. In our simulations, which use $O(10^6)$ particles, we expect this wobbling to be on the order of a few parsecs and less than a ${\rm km\,s}^{-1}$. \citet{lau2021} showed that Poisson noise can excite macroscopic modes leading to random motions larger than this by an order of magnitude. In fact, we find that in the IH simulation, the disc wobbles by $50-100\,{\rm pc}$.

\begin{figure}
	\includegraphics[width=\columnwidth]{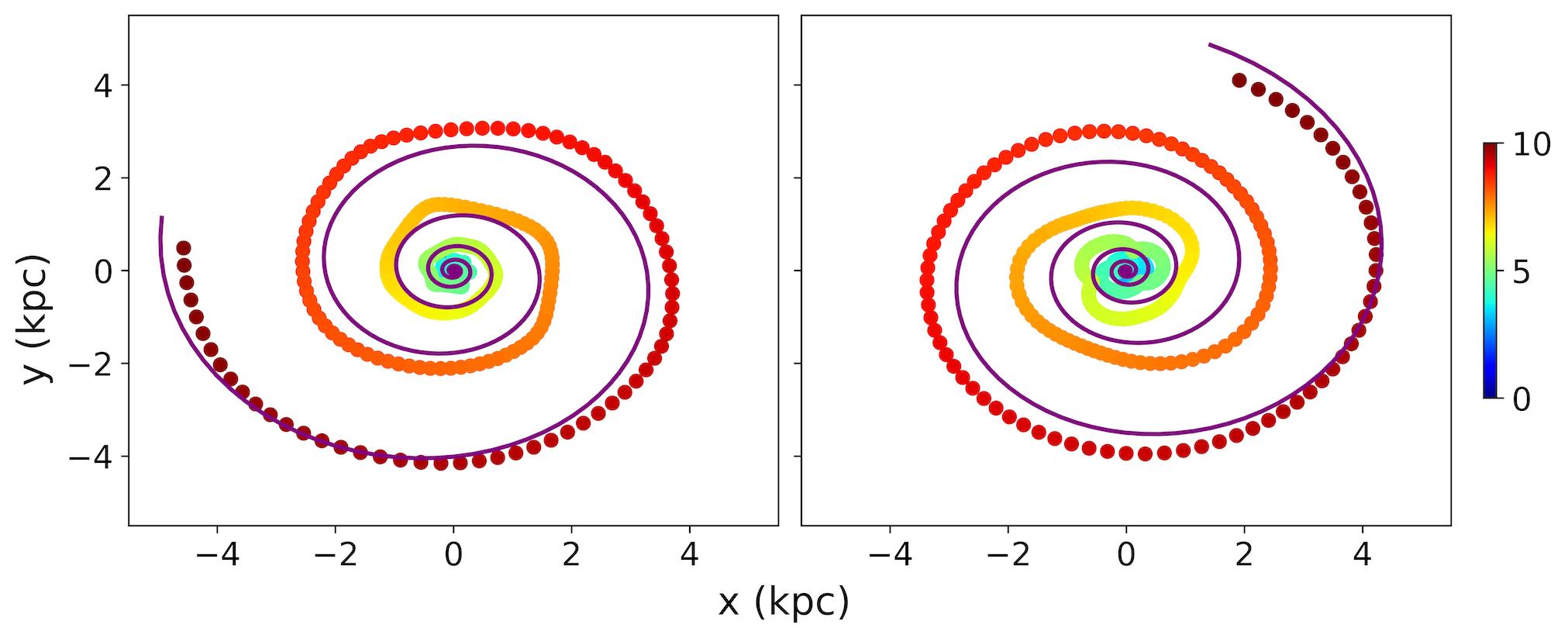}
    \caption{The center of mass of the stellar disk over the duration of the simulation for the RH simulation (left) and PH simulation (right). The points along the center-of-mass orbits are color-coded by time in Gyr as indicated by the color bar. The purple line is the prediction of our model until t = 10 Gyr.}
    \label{fig:centerofmass}
\end{figure}

Unexpectedly, we find that in the PH and RH simulations, the disc as well as the halo cusp spirals outward from the centre of mass of the system as a whole by several kiloparsecs. This effect is clearly seen in Fig. \ref{fig:centerofmass} where we plot the position of the center of mass of the disc in the $x-y$ plane color-coded by time.

\begin{figure}
	\includegraphics[width=\columnwidth]{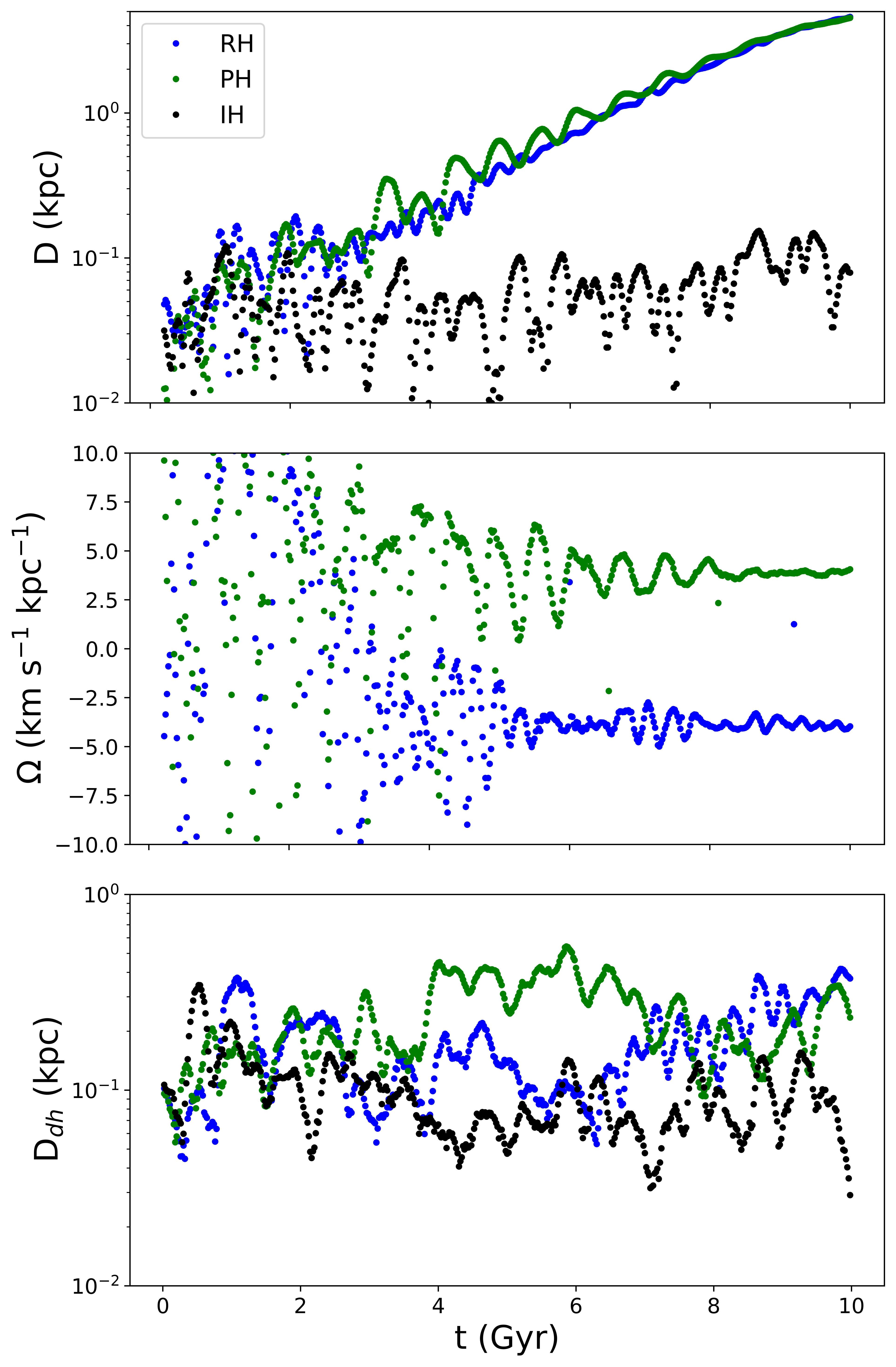}
    \caption{Radial distance $D$ (top panel) and angular velocity $\Omega$ (middle panel) of the disc center of mass with respect to the origin of the simulation coordinate system. Black, green, and blue are for IH, PH, and RH cases, respectively. The angular velocity for the IH case is omitted. The bottom panel shows the distance between the center of mass of the disc and the cusp of the halo.}
    \label{fig:cmdistance}
\end{figure}

In Fig.~\ref{fig:cmdistance} we show the distance $D$ between the center of mass of the disc and the origin of the simulation coordinate system. Though the discs in the RH and PH simulations spiral in opposite directions, as seen in Fig.~\ref{fig:centerofmass}, the time-dependence of $D$ is remarkably similar in the two cases. The growth is approximately exponential from the start of the simulations, when $D\simeq 30\,{\rm pc}$ to $D\simeq 3\,{\rm kpc}$ at $8\,{\rm Gyr}$. This gives an efolding time for $D$ of $\tau\simeq 1.7\,{\rm Gyr}$. At later times, the growth is sub-exponential suggesting that the instability has saturated. By contrast, the disc stays within $\sim 100\,{\rm pc}$ in the IH simulation. This is somewhat larger than what is expected from purely Poisson noise, but, as discussed in Section 3, the bar and warp include odd-$m$ components.

In Fig.~\ref{fig:cmdistance} we also show the angular frequency $\Omega$ of the disc centre. As can been seen from this figure and also Fig.~\ref{fig:centerofmass}, $\Omega\simeq \pm 5\,{\rm kpc\,km\,s}^{-1}\simeq 5\,{\rm Gyr}^{-1}$, which is considerably lower than the pattern speed of the bar in all of our simulations except for the final few gigayears of the RH case when the pattern speed of the bar has slowed considerably. Finally, we show the distance between the disc and halo cusp. The position of the halo cusp is estimated by first making a two-dimensional histogram of the number of halo particles within $2\,{\rm kpc}$ of the $xy$-plane and identifying the bin with the largest number of particles. A simple quadratic fit of the number counts in a $5\times 5$ grid surrounding the peak bin then provides the desired approximate position of the true density peak in the $xy$-plane. We then calculate the distance $D_{dh}$ between the cusp and disc centre of mass as a function of time. As we see, the cusp of the halo tracks the disc as it spirals outward though the disc and cusp can be separated by as much as $500\,{\rm pc}$. The separation of a stellar disc and halo cusp of a few hundred parsecs has been seen in the hydrodynamic simulations of Milky Way-like galaxies by \citet{kuhlen2013}. The effect could be important for understanding the dynamics of the Galactic centre as well as dark matter annihilation signals. 

\begin{figure}
\includegraphics[width=\columnwidth]{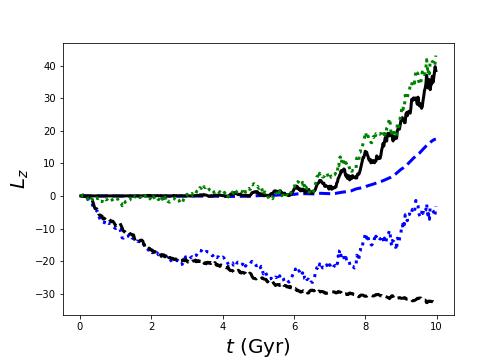}
\caption{Spin and orbital angular momentum in the disc as a function of time for the PH simulation. The solid black curve shows ${\bf L}_{\rm orb}$ while the dashed black curve shows ${\bf L}_{\rm spin}$ minus its initial value. The dotted and dashed blue curves show the angular momentum imparted to the disc by the inner and outer halo, respectively. The dotted green curve is the sum of the terms on the right-hand side of equation \ref{eq:RHI_AM}. The units for $L_z$ are $10^{11}\,M_\odot {\rm kpc} \,{\rm km\,s}^{-1}$. Conservation of angular momentum predicts that it should match the solid black curve (left-hand side of the same equation).}
    \label{fig:RHI_AM}
\end{figure}

The motion of the disc in the RH and PH simulations can be understood in terms of a transfer of angular momentum between the disc and halo. We can write the positions and velocities of the disc particles as ${\bf r}_n = \mathbfcal{R} + \bf{s}_n$ and ${\bf v} = \mathbfcal{V} + \bf{u}_n$ where $\mathbfcal{R}$ and $\mathbfcal{V}$ are the position and velocity of the disc center of mass. The angular momentum of the disc then splits into spin and orbital contributions:
\begin{equation}
{\bf L} =
M_d \mathbfcal{R}\times  \mathbfcal{V} + 
\sum_n m_n{\bf s}_{n} \times {\bf u}_{n} 
= {\bf L}_{\rm orb} + {\bf L}_{\rm spin}~.
\label{eq:spinorbit}
\end{equation}
The evolution of the disc orbital angular momentum is then given by the equation
\begin{equation}
{\bf L}_{\rm orb}(t) = -({\bf L}_{\rm spin}(t)-{\bf L}_0) + \int_0^t
dt' \boldsymbol{tau}_{\rm halo}(t')
\label{eq:RHI_AM}
\end{equation}
where ${\bf L}_0$ is the initial spin angular momentum of the disc and $\tau_{\rm halo}$ is the torque acting on the disc due to the halo. Fig.\ref{fig:RHI_AM} provides a graphical representation of this equation for the RH simulation. We calculate the torque of the halo on the disc for a sequence of snapshots spaced by $20\,{\rm Myr}$ and carry out the integral in equation \ref{eq:RHI_AM} using the trapezoidal rule. The forces that are required for the torque are computed using \textsc{pytreegrav} \citet{pytreegrav2021}, a \textsc{python} code based on the tree method of \citet{barnes1986} that allows us to separate out the forces from the halo and disc. We compute the separate contributions from the inner and outer halo where the inner (outer) halo includes particles inside (outside) a sphere of radius $25\,{\rm kpc}$ centred on the origin of the simulation volume. We choose $25\,{\rm kpc}$ as this is roughly the extent of the disc. Though our distinction between inner and outer halo is rather ad hoc it allows us to roughly see which regions of the halo are torquing the disc. The black curve in Fig.\ref{fig:RHI_AM} is the orbital angular momentum of the disc and rises exponentially as expected from the previous figures. The dotted green curve is the sum of the terms on the right-hand side of equation \ref{eq:RHI_AM} and tracks ${\bf L}_{\rm orb}$, as one expects from conservation of angular momentum. We attribute the difference between the solid black and dotted green curves as due to the crude approximation of the integral in equation \ref{eq:RHI_AM}. The dashed black curve shows the spin angular momentum of the disc minus its initial value. As discussed above the disc loses spin angular momentum over the course of the simulation. Here it is clear that this is mainly due to the torque from the inner halo. However, by $5\,{\rm Gyr}$ the disc is clearly gaining orbital angular momentum, evidently due to torques from both the inner and outer halos. Though the halo causes the disc and bar to spin down, it transfers its own spin angular momentum to orbital angular momentum to the disc as a whole, causing the disc centre to spiral outward.

\begin{figure}
	\includegraphics[width=\columnwidth]{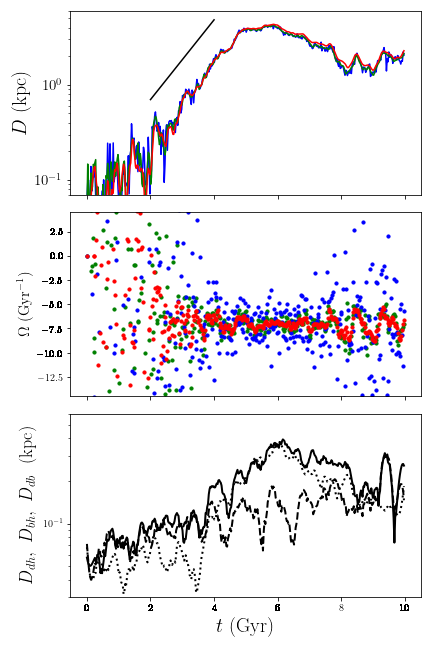}
    \caption{Displacement of the disc, bulge, and halo cusp as a function of time for a simulation where the disc is stable against the formation of a strong bar. The top panel shows the distance of the disc (red), bulge (green), and halo cusp (blue) from the origin of the simulation. The solid black line shows the growth rate derived from the model fit (see text). The middle panel shows the angular frequency of these systems. The bottom panel shows the distances between the disc and bulge ($D_{db}$, dotted curve), the disc and halo cusp ($D_{dh}$, solid curve) and bulge and cusp ($D_{bh}$, dashed curve)}.
    \label{fig:DBH-Displacement.png}
\end{figure}

To further investigate the RHI, we followed the evolution of a very different model, namely one with a disc, bulge, and rotating halo that doesn't form a bar. In this case, we chose a less massive and warmer disc than our previous models ($M_d \simeq 3.5\times 10^{10}\,M_\odot$ and $Q=1.5$ for $R=2.2R_d$). For brevity, we only present results in the retrograde case. The top panel shows the distance of the disc, bulge, and halo cusp from the origin of the simulation. Evidently, the three components track each other fairly well. The exponential growth rate is $\simeq 0.96\,{\rm Gyr}^{-1}$ as compared with $0.59\,{\rm Gyr}^{-1}$ in the disc-halo models discussed above. Likewise, the angular frequency is higher by a similar factor ($|\Omega|\simeq 7\,{\rm Gyr}^{-1}$ as compared with $5\,{\rm Gyr}^{-1}$). The instability saturates at about $5\,{\rm Gyr}$, when the the distance of the disc to the origin of the simulation is $\sim 3\,{\rm kpc}$. Though saturation of the instability occurs earlier than in the disc-halo models, the maximum excursion appears to be about the same. In the bottom panel we show the distances between the disc and halo cusp, $D_{dh}$, the disc and bulge, $D_{db}$ and the bulge and halo, $D_{bh}$ as functions of time. As in Figure 11, we define the positions of the disc and bulge by their centre of mass and the position of the halo by the position of the cusp as described above. We see that the tightest coupling is between the bulge and halo cusp while the disc wanders from the bulge and cusp by a few hundred parsecs between $5-7\,{\rm Gyr}$.

We propose that the motion of the disc seen in Figs. \ref{fig:centerofmass} and \ref{fig:cmdistance} is the result of a linear instability. Consider the following heuristic equation of motion for the disc:
\begin{equation}
  \frac{d{\bf V}}{dt} = -\alpha^2 {\bf R}
  -\beta \left ({\bf V} -  \omega \hat{\bf z}\times {\bf R} \right )
\end{equation}
where ${\bf R}=(X,Y)$ and ${\bf V}=(V_x,V_y)$ are the position and velocity of the center of mass of the disc relative to the center of mass of the system. In this simple model the parameters $\alpha$, $\beta$, and $\omega$ are constants of dimensions of inverse time. The first term accounts for the gravitational attraction of the disc to the center of mass of the system. On dimensional grounds we expect $\alpha^2\sim 4\pi G\rho$ where $\rho$ is an effective mean density of the halo in the region of the disc. The second term accounts for an effective drag on the system due to collective effects in the halo as the disc moves through it. The $\beta {\bf V}$-term is reminiscent of the Chandrasekhar formula for dynamical friction. Again, on dimensional grounds, we expect $\beta\simeq 4\pi G\rho \left (GM/\sigma_h^3\right )f(V/\sigma_h)$ where $\sigma_h$ is the velocity dispersion of the halo in the disc region and $f$ is a dimensionless function that approaches $\sqrt{2/9\pi}$ as $V\to 0$ and approached zero for $V\to \infty$. The term $\omega \hat{\bf z}\times {\bf R}$ modifies the velocity dependent term to incorporate bulk rotational motion of the halo particles. The parameter $\omega$ can be positive or negative; its sign determines whether the halo is prograde or retrograde.

The equations for $X$ and $Y$ can be combined to yield a fourth-order linear homogeneous ordinary differential equation. With the ansatz $(X,Y) = (X_0,Y_0)e^{i\mu t}$ we obtain the equation
\begin{equation}
  ( \mu^2 -i\beta\mu -\alpha^2)^2 = -\beta^2\omega^2
  \label{eq:RHImodel}
\end{equation}
This quartic equation has four solutions
\begin{align}
  \mu & = \frac{i\beta}{2} \pm\left (\alpha^2 - \frac{\beta^2}{4} \pm i\beta\omega\right )^{1/2}\\
  & = i\left (\frac{\beta}{2} \pm \gamma \sin{\varphi}\right ) \pm \gamma\cos{\varphi}
\end{align}
where 
\begin{equation}
\gamma = \left (\left (\alpha^2 -\frac{\beta^2}{4}\right )^2 + \beta^2\omega^2\right )^{1/4},
\end{equation}
\begin{equation}
\tan{(2\varphi)} = \frac{\beta\omega}
{\left (\alpha^2-\frac{\beta^2}{4}\right)},
\end{equation}
and all four sign combinations are allowed. The most general solution for $X(t)$ has both growing and decaying solutions,
\begin{align}
  X(t)  = A_g &e^{(\gamma\sin{\varphi-\beta/2})t}\cos{(\gamma t\cos{\varphi} +\vartheta_g)}\nonumber\\ & + 
  A_de^{-(\gamma\sin{\varphi}+\beta/2)t}\cos{(\gamma t\cos{\varphi } + \vartheta_d)}
\end{align}
where the four constants $A_{g,d}$ and $\vartheta_{g,d}$ are determined by initial conditions. At late times, the solution is dominated by the growing solution
\begin{equation}
  X(t) = A_g e^{\lambda t} \cos{(\Omega t + \vartheta_g)}
\end{equation}
where $\lambda \equiv \gamma \sin{\varphi} - \beta/2$ and $\Omega \equiv
\gamma\cos{\varphi}$. Note that if we ignore the dynamical friction term and set $\beta=0$, we find $\lambda=0$ and $\Omega = \alpha$. In this case, the disc orbits about the center of mass along a closed ellipse. Of course, in a more realistic model, the first term on the right hand side of equation \ref{eq:RHImodel} would include non-linearities and the orbit would follow a rosette pattern \citep{binney2008}. In the case where we include the friction term but not halo rotation, the disc would spiral inward. This case is reminiscent of the sinking satellite problem considered, for example, in \citet{lin1983, white1983, duncan1983, velazquez1999}. As discussed in those works and numerous other investigations, there is considerable debate regarding the applicability of the Chandrasekhar formula to the decay of satellite orbits. The application of the Chandrasekhar formula in our case is also suspect since the halo will adjust to the disc. Indeed we find that the halo cusp moves in sync with the disc. Nevertheless, the basic idea of a friction term that is proportional to ${\bf V}$ seems to capture the physics through our simple model even if the magnitude of the effect is treated as a free parameter.

With halo rotation ($\omega\ne 0$), we have the possibility for the disc to spiral outward. Physically, the disc is being swept up in the cyclonic dark matter wind. So long as $\gamma\sin{\phi}-\beta/2>0$ and the approximations that went into equation \ref{eq:RHImodel} hold the distance between the disc and system centre of mass will grow. A fit of the disc orbit in Fig.~\ref{fig:centerofmass} to this model yields $\lambda \simeq 0.55\,{\rm Gyr}^{-1}$ and $\Omega\simeq 4.1\,{\rm Gyr}^{-1}$, or equivalently, an e-folding time for the distance between the disc/cusp of $1.8\,{\rm Gyr}$ and an orbital period of $1.5\,{\rm Gyr}$.

%A theoretical calculation of the coupling between the disc/cusp and the rotating halo is beyond the scope of this paper. However, we can obtain the following order-of-magnitude estimate for $\beta$. From the standard dynamical friction formula \citep{chandra1943, binney2008}, we have
%\begin{equation}
%\beta \sim 4\pi G\rho_h(R) \left (GM/V_M\right )^3
%\end{equation}
%where $\rho_h(R)$ is the density of the halo, $M$ is the mass of the cusp/disc that is being swept outward by dynamical friction, and $V_M$ is the velocity of $M$ relative to the rotating halo. There are, of course, numerical factors that come from the Coulomb logarithm and the velocity integral. If we evaluate $\rho_h$ at $R=R_d$, set $M=M_d$, and $V_M = 220\,{\rm km\,s}^{-1}$, the circular speed of the disc, we find $\beta\sim 50\,{\rm Gyr}^{-1}$, which is within a factor of a few of the value used in Fig.~\ref{fig:centerofmass}.

\section{Discussion and Conclusion}
\label{sec:conclusion}

The simulations in this paper were designed to explore the effects of halo rotation on the formation and evolution of bars and warps in disc galaxies. To that end, we simulated identical discs in four haloes that were characterized by very different DFs. In each case, the disc formed a strong bar, which subsequently went through a sequence of buckling events. In our live-halo simulations the formation of a strong bar in the disc was accompanied by the formation of a shadow bar in the dark halo. In addition, the discs embedded in live halos developed warps with amplitudes that ranged from $100\,{\rm pc}$ to $300\,{\rm pc}$. 

A recurring theme in this work is that the formation and evolution of bars and warps depends on the orbits of the halo particles. For example, the bar forms first but buckles last in the PH case and vice versa in the RH case. Evidently, the presence of co-rotating halo particles in the PH simulation facilitates the transfer of angular momentum to the disc thereby helping to drive the formation of the bar. However, the co-rotating halo is more susceptible to the formation of a shadow bar, which slows the growth and spin-down of the bar. For this reason, what we refer to as a late-stage buckling event is delayed. The results are consistent with those of earlier studies such as those by \cite{sellwood2016bar, collier2021coupling} and \citet{kataria2022effects}.

After the bar instability occurs it is the bar in the RH simulation that grows the most as it slows down during secular evolution. On the other hand, the length of the PH bar remains constant. The results of \citet{saha2013} are seemingly in conflict with our findings but it is important to point out that the secular evolution of retrograde and prograde halo bars was not addressed at length in this study. More recent work by \citet{collier2019stellar} showed that after the bar instability occurs, orbit reversals in a retrograde halo due to the bar will slow the bar causing it to grow. This effect is not as prominent in a prograde halo since the bar traps dark matter particles in a dark bar, resulting in a stagnant bar that does not slow significantly.

We have identified two distinct modes for buckling. The buckling events referred to in this work as late stage are well-studied. The bar bends in opposite directions on either side of the center of the disc leading to a 'V'-shape pattern in the surface density, when projected along the minor axis of the bar. The bulk motions along the bar can be viewed as a bending or circulation in opposite directions on either side of the center of the disc \citep{li2022stellar, li2023}.

In our live halo simulations, the bar undergoes oscillations perpendicular to the plane almost immediately after forming. The vertical displacement profile for these early-stage buckling events have a 'tilde'-shaped pattern in surface density when viewed edge on. The pattern is generated by having the bar bend clockwise in the center of the bar and counterclockwise in the outer parts, or vice versa. Interestingly enough, in our live halo simulations this type of buckling event occurs almost immediately after the bar forms but after a period of about $1\,{\rm Gyr}$ in the case where the halo is static. Moreover, in the SH case, the bar never experiences the normal or late-stage buckling event seen so prominently in the live-halo simulations.

The most novel result in this work is the discovery of an instability in the motion of the disc and halo cusp that arises in our rotating halo simulations. Essentially, the disc and halo cusp are swept outward by the rotating halo particles. The RHI can be explained by a heuristic formula based on the effects of dynamical friction between the halo and the disc. As striking as the instability is, it is worth stressing that our simulations start from highly idealized conditions. In future studies we will investigate whether the instability acts on discs and haloes that form from cosmological initial conditions.

\section*{Acknowledgements}

We thank Martin Weinberg and Lucio Mayer for useful conversations. This work was supported by a Discovery Grant through the Natural Sciences and Engineering Research Council of Canada.

%%%%%%%%%%%%%%%%%%%%%%%%%%%%%%%%%%%%%%%%%%%%%%%%%%
\section*{Data Availability}

The data underlying this article were generated by numerical calculations using original \textsc{Python} code written by the author. The code incorporated routines from \textsc{NumPy} \citep{harris2020} and \textsc{SciPy} \citep{virtanen2020}. The data for the figures and the code will be shared on reasonable request to the author.



\bibliographystyle{mnras}
\bibliography{BarHaloInteraction} % if your bibtex file is called example.bib

%\begin{thebibliography}{99}
%\end{thebibliography}

%%%%%%%%%%%%%%%%%%%%%%%%%%%%%%%%%%%%%%%%%%%%%%%%%%

%%%%%%%%%%%%%%%%% APPENDICES %%%%%%%%%%%%%%%%%%%%%

% Don't change these lines
\bsp	% typesetting comment
\label{lastpage}
\end{document}